\documentclass[twocolumn,showpacs,preprintnumbers,amsmath,amssymb,floatfix,superscriptaddress]{revtex4}
\usepackage{graphicx}% Include figure files
\usepackage{dcolumn}% Align table columns on decimal point
\usepackage{bm}% bold math

\begin{document}

\title{$K/\pi$ Fluctuations at Relativistic Energies}% Force line breaks with \\

\affiliation{Argonne National Laboratory, Argonne, Illinois 60439, USA}
\affiliation{University of Birmingham, Birmingham, United Kingdom}
\affiliation{Brookhaven National Laboratory, Upton, New York 11973, USA}
\affiliation{University of California, Berkeley, California 94720, USA}
\affiliation{University of California, Davis, California 95616, USA}
\affiliation{University of California, Los Angeles, California 90095, USA}
\affiliation{Universidade Estadual de Campinas, Sao Paulo, Brazil}
\affiliation{University of Illinois at Chicago, Chicago, Illinois 60607, USA}
\affiliation{Creighton University, Omaha, Nebraska 68178, USA}
\affiliation{Nuclear Physics Institute AS CR, 250 68 \v{R}e\v{z}/Prague, Czech Republic}
\affiliation{Laboratory for High Energy (JINR), Dubna, Russia}
\affiliation{Particle Physics Laboratory (JINR), Dubna, Russia}
\affiliation{Institute of Physics, Bhubaneswar 751005, India}
\affiliation{Indian Institute of Technology, Mumbai, India}
\affiliation{Indiana University, Bloomington, Indiana 47408, USA}
\affiliation{Institut de Recherches Subatomiques, Strasbourg, France}
\affiliation{University of Jammu, Jammu 180001, India}
\affiliation{Kent State University, Kent, Ohio 44242, USA}
\affiliation{University of Kentucky, Lexington, Kentucky, 40506-0055, USA}
\affiliation{Institute of Modern Physics, Lanzhou, China}
\affiliation{Lawrence Berkeley National Laboratory, Berkeley, California 94720, USA}
\affiliation{Massachusetts Institute of Technology, Cambridge, MA 02139-4307, USA}
\affiliation{Max-Planck-Institut f\"ur Physik, Munich, Germany}
\affiliation{Michigan State University, East Lansing, Michigan 48824, USA}
\affiliation{Moscow Engineering Physics Institute, Moscow Russia}
\affiliation{City College of New York, New York City, New York 10031, USA}
\affiliation{NIKHEF and Utrecht University, Amsterdam, The Netherlands}
\affiliation{Ohio State University, Columbus, Ohio 43210, USA}
\affiliation{Old Dominion University, Norfolk, VA, 23529, USA}
\affiliation{Panjab University, Chandigarh 160014, India}
\affiliation{Pennsylvania State University, University Park, Pennsylvania 16802, USA}
\affiliation{Institute of High Energy Physics, Protvino, Russia}
\affiliation{Purdue University, West Lafayette, Indiana 47907, USA}
\affiliation{Pusan National University, Pusan, Republic of Korea}
\affiliation{University of Rajasthan, Jaipur 302004, India}
\affiliation{Rice University, Houston, Texas 77251, USA}
\affiliation{Universidade de Sao Paulo, Sao Paulo, Brazil}
\affiliation{University of Science \& Technology of China, Hefei 230026, China}
\affiliation{Shandong University, Jinan, Shandong 250100, China}
\affiliation{Shanghai Institute of Applied Physics, Shanghai 201800, China}
\affiliation{SUBATECH, Nantes, France}
\affiliation{Texas A\&M University, College Station, Texas 77843, USA}
\affiliation{University of Texas, Austin, Texas 78712, USA}
\affiliation{Tsinghua University, Beijing 100084, China}
\affiliation{United States Naval Academy, Annapolis, MD 21402, USA}
\affiliation{Valparaiso University, Valparaiso, Indiana 46383, USA}
\affiliation{Variable Energy Cyclotron Centre, Kolkata 700064, India}
\affiliation{Warsaw University of Technology, Warsaw, Poland}
\affiliation{University of Washington, Seattle, Washington 98195, USA}
\affiliation{Wayne State University, Detroit, Michigan 48201, USA}
\affiliation{Institute of Particle Physics, CCNU (HZNU), Wuhan 430079, China}
\affiliation{Yale University, New Haven, Connecticut 06520, USA}
\affiliation{University of Zagreb, Zagreb, HR-10002, Croatia}

\author{B.~I.~Abelev}\affiliation{University of Illinois at Chicago, Chicago, Illinois 60607, USA}
\author{M.~M.~Aggarwal}\affiliation{Panjab University, Chandigarh 160014, India}
\author{Z.~Ahammed}\affiliation{Variable Energy Cyclotron Centre, Kolkata 700064, India}
\author{B.~D.~Anderson}\affiliation{Kent State University, Kent, Ohio 44242, USA}
\author{D.~Arkhipkin}\affiliation{Particle Physics Laboratory (JINR), Dubna, Russia}
\author{G.~S.~Averichev}\affiliation{Laboratory for High Energy (JINR), Dubna, Russia}
\author{J.~Balewski}\affiliation{Massachusetts Institute of Technology, Cambridge, MA 02139-4307, USA}
\author{O.~Barannikova}\affiliation{University of Illinois at Chicago, Chicago, Illinois 60607, USA}
\author{L.~S.~Barnby}\affiliation{University of Birmingham, Birmingham, United Kingdom}
\author{J.~Baudot}\affiliation{Institut de Recherches Subatomiques, Strasbourg, France}
% ...

\collaboration{STAR Collaboration}\noaffiliation

\date{\today}% It is always \today, today,
             %  but any date may be explicitly specified

\begin{abstract}
We report results for $K/\pi$ fluctuations from Au+Au collisions at $\sqrt{s_{NN}}$ = 19.6, 62.4, 130, and 200 GeV using the STAR detector at the Relativistic Heavy Ion Collider. Our results for $K/\pi$ fluctuations in central collisions show little dependence on the incident energies studied and are on the same order as results observed by NA49 at the Super Proton Synchrotron in central Pb+Pb collisions at $\sqrt{s_{NN}}$ = 12.3 and 17.3 GeV. We also report results for the collision centrality dependence of $K/\pi$ fluctuations as well as results for $K^{+}/\pi^{+}$, $K^{-}/\pi^{-}$, $K^{+}/\pi^{-}$, and $K^{-}/\pi^{+}$ fluctuations. We observe that the $K/\pi$ fluctuations scale with the multiplicity density, $dN/d\eta$, rather than the number of participating nucleons.  
\end{abstract}

\pacs{25.75.-q, 24.60.Ky}

\keywords{fluctuations, strangeness, $K/\pi$ ratios, relativistic heavy ions, STAR, RHIC} 
  
\maketitle

Strangeness enhancement has been predicted to be one of the important signatures of the formation of the quark gluon plasma (QGP) \cite{early_strangeness,strangeness,koch1,koch2,gorenstein1}. Recently a maximum in the ratio of the yields of $K^{+}$ and $\pi^{+}$, $K^{+}/\pi^{+}$, has been observed in central Pb+Pb collisions near $\sqrt{s_{NN}}$ = 7 GeV \cite{NA49_horn_step}.   Dynamical fluctuations in the event-by-event $K/\pi$ ratio in central Pb+Pb collisions at energies near $\sqrt{s_{NN}}$ = 7 GeV are larger than those predicted by the transport model UrQMD using the observable $\sigma_{\rm dyn}$ \cite{NA49 fluctuations 1}. The $K/\pi$ ratio is defined as $K/\pi \equiv (N_{K^{+}} + N_{K^{-}})/(N_{\pi^{+}} + N_{\pi^{-}})$, where $(N_{K^{+}} + N_{K^{-}})$ is the number of charged kaons in one event and $(N_{\pi^{+}} + N_{\pi^{-}})$ is the number of charged pions in the same event.  The observable $\sigma_{\rm dyn}$ \cite{NA49 fluctuations 2} is defined as
\begin{equation}
\label{ }
\sigma_{\rm dyn}=sign\left( \sigma_{\rm data}^2 - \sigma_{\rm mixed}^2 \right)\sqrt{\left| \sigma_{\rm data}^2 - \sigma_{\rm mixed}^2 \right|}
\end{equation}
where $\sigma_{\rm data}$ is the relative width (standard deviation divided by the mean) of the $K/\pi$ distribution for the data and $\sigma_{\rm mixed}$ is the relative width of the $K/\pi$ distribution for mixed  events. These observations have generated speculation that a phase transition from hadronic matter to quark-gluon matter may be taking place at incident energies around $\sqrt{s_{NN}}$ = 7 GeV \cite{NA49_horn_step}.  The study of dynamic fluctuations in the event-by-event $K/\pi$ ratio may produce information concerning QCD phase transitions such as the order of the transitions and the location of the transitions, and may lead to the observation of the critical point of QCD \cite{tricritical,criticalpoint}.

In this paper, we report results for dynamic fluctuations of the $K/\pi$ ratio in central Au+Au collisions at $\sqrt{s_{NN}}$ = 19.6, 62.4, 130, and 200 GeV using the quantity $\sigma_{\rm dyn}$ \cite{NA49 fluctuations 2}.  These results are compared with the results of NA49 for dynamical $K/\pi$ fluctuations in central Pb+Pb collisions \cite{NA49 fluctuations 1}. To study the collision centrality dependence of $K/\pi$ fluctuations, we propose a new variable, $\nu_{{\rm dyn},K\pi}$, which quantifies the deviation of the fluctuations in the number of pions and kaons from that expected from Poisson statistics.  This variable is defined as
	\begin{eqnarray}
\nu _{{\rm dyn},K\pi }^{}  = \frac{{\left\langle {N_K \left( {N_K  - 1} \right)} \right\rangle }}{{\left\langle {N_K } \right\rangle ^2 }} + \frac{{\left\langle {N_\pi  \left( {N_\pi   - 1} \right)} \right\rangle }}{{\left\langle {N_\pi  } \right\rangle ^2 }} \\ \nonumber - 2\frac{{\left\langle {N_K N_\pi  } \right\rangle }}{{\left\langle {N_K } \right\rangle \left\langle {N_\pi  } \right\rangle }}
	\end{eqnarray}
where $N_{\pi}$ is the number of pions in each event and $N_{K}$ is the number of kaons in each event.   The properties of Eq. 2 are discussed at length in Ref. \cite{nudyn}.   Negative values of $\nu_{{\rm dyn},K\pi}$ imply that the third term in Eq. 2 involving $K-\pi$ correlations dominates, while positive values of $\nu_{{\rm dyn},K\pi}$ imply that the first two terms involving the joint correlations $K-K$ and $\pi-\pi$ dominate.  We present results for the collision centrality dependence of $\nu_{{\rm dyn},K\pi}$ for Au+Au collisions at 62.4 and 200 GeV.   To gain insight concerning the origins of these $K/\pi$ fluctuations \cite{chemical_fluctuations}, we also present the collision centrality dependence of  $\nu_{{\rm dyn},K\pi}$ for  $K^{+}/\pi^{+}$, $K^{-}/\pi^{-}$, $K^{+}/\pi^{-}$, and $K^{-}/\pi^{+}$.  

Au+Au collisions were studied at $\sqrt{s_{NN}}$ = 19.6, 62.4, 130, and 200 GeV using the STAR detector at the Relativistic Heavy Ion Collider (RHIC).   All data presented here were taken using a minimum bias trigger.  Events accepted took place within $\pm 15$ cm of the center of the STAR detector in the beam direction.  Collision centrality was determined using the number of charged tracks within $|\eta| < 0.5$.  Nine centrality bins were used corresponding to 0-5\% (most central), 5-10\%, 10-20\%, 20-30\%, 30-40\%, 40-50\%, 50-60\%, 60-70\%, and 70-80\% (most peripheral) of the reaction cross section.  To be able to plot our results versus $dN/d\eta$, we associate
fully corrected values for $dN/d\eta$ from previously published work with each collision centrality bin 
\cite{star_dNdeta}. 
For the 19.6 GeV and 130 GeV data sets, only results from the most central bin are presented.  All tracks were required to have originated within 3 cm of the measured event vertex.
Only charged particle tracks having more than 15 space points along the trajectory were accepted. The
ratio of reconstructed points to possible points along the
track was required to be greater than 0.52 to avoid split tracks.  Charged pions and charged kaons were identified using the specific energy loss, $dE/dx$, along the track and the momentum, $p$, of the track.

Charged pions and kaons were selected with transverse momentum $0.2 < p_{t} < 0.6$ GeV/c and pseudorapidity $|\eta| < 1.0$. Particle identification was accomplished by selecting particles whose specific energy losses were within two standard deviations of the energy loss predictions for a given particle type and momentum.  Particle identification for pions (kaons) also included a condition that the specific energy loss should be more than two standard deviations away from the loss predicted for a kaon (pion).
In addition, electrons were excluded from the analysis for all cases.  Particles were excluded as electrons if the specific energy losses were within one standard deviation of the energy loss predictions for electrons.  We identify $\sim$90\% of the pions in our acceptance.  We identify $\sim$50\% of the kaons at $p_{t}$ = 0.2 GeV/c and $\sim$75\% of the kaons at $p_{t}$ = 0.6 GeV/c in our acceptance.  We calculate that the fraction of pions resulting from misidentified kaons is negligible while the fraction of kaons resulting from misidentified pions is 6.5\%.  The electron cut did not affect the pions significantly, but excluded 25\% of the kaons for the 200 GeV Au+Au case and 35\% of the kaons for the 62.4 GeV Au+Au system.  The remaining electron contamination is negligible.

\begin{figure}
	\centering
	\includegraphics[width=3.0 in]{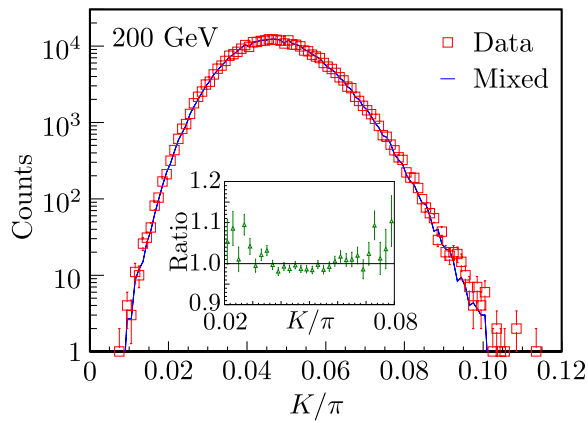}
	\caption{\label{fig:fig1}(Color online) The event-by-event $K/\pi$ ratio for 200 GeV Au+Au central collisions (0-5\%) compared with the same quantity calculated from mixed events.  The inset shows the ratio of the distribution from real events to that from mixed events.  The errors shown are statistical.}
\end{figure}

In Fig.~\ref{fig:fig1}, we show the distribution of the event-by-event $K/\pi$ ratio for central Au+Au collisions (0-5\%) at $\sqrt{s_{NN}}$ = 200 GeV compared with the same quantity from mixed events.  Mixed events were created by taking one track from different events to produce new events with the same multiplicity that have no correlations among particles in those events.  Mixed events were produced using ten bins in collision centrality and five bins in event vertex position.  The distribution for the data is wider than the distribution for mixed events.  Similar results were obtained at the other three incident energies; 19.6, 62.4, and 130 GeV.  The same analysis techniques are applied to the mixed events as are applied to the data.

\begin{figure}
	\centering
	\includegraphics[width=3.0 in]{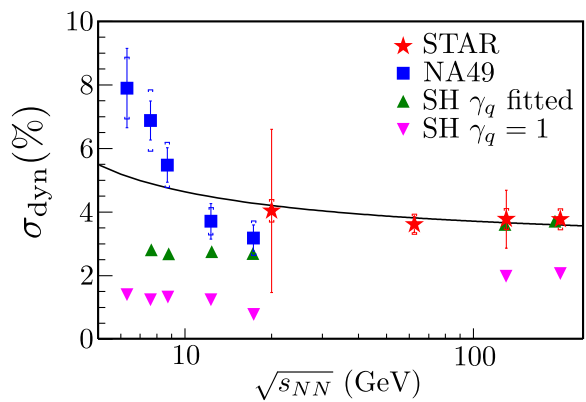}
	\caption{\label{fig:fig2}(Color online) Measured dynamical $K/\pi$ fluctuations in terms of $\sigma_{\rm dyn}$ for central collisions (0 - 5\%) of 19.6, 62.4, 130, and 200 GeV Au+Au compared with the central collisions (0 - 3.5\%) of Pb+Pb from NA49 \cite{NA49 fluctuations 1} and the statistical hadronization (SH) model of Ref. \cite{torrieri}.  The solid line represents the relationship of the incident energy dependence of $\sigma_{\rm dyn}$ in central collisions to the collision centrality dependence of $\nu_{{\rm dyn},K\pi}$ at higher energies.  Both statistical (vertical line with horizontal bar) and systematic (no vertical line) error bars are shown for the experimental data.}
\end{figure}

The results for $\sigma_{\rm dyn}$ from central Au+Au collisions at 19.6, 62.4, 130, and 200 GeV are shown in Fig.~\ref{fig:fig2} along with similar results for  central Pb+Pb collisions at $\sqrt{s_{NN}}$ = 6.3, 7.6, 8.8, 12.3, and 17.3 GeV from the NA49 Collaboration \cite{NA49 fluctuations 1}.  Statistical and systematic errors are shown for both the NA49 results and the STAR results.  The systematic errors for $\sigma_{\rm dyn}$ are discussed in the presentation of the results for $\nu_{{\rm dyn},K\pi}$. 

In Fig.~\ref{fig:fig2}, we find that the NA49 results show a strong incident energy dependence while the STAR results show little dependence on the incident energy.  The STAR results are consistent with the highest energy NA49 result, although the statistical error bar for the 19.6 GeV case is large.  In this figure, we compare the statistical hadronization model results of Torrieri \cite{torrieri} to the experimental data.   We see that when the light quark phase space occupancy, $\gamma_{q}$, is one, corresponding to equilibrium, the calculations underestimate the experimental results at all energies.  When $\gamma_{q}$ is varied to reproduce the excitation function of $K^{+}/\pi^{+}$ yield ratios and the excitation function of temperature versus chemical equilibrium over the SPS and RHIC energy ranges \cite{torrieri,rafelski} , the statistical hadronization model  correctly predicts the dynamical fluctuations at the higher energies but under-predicts the NA49 data at the lower energies, supporting the conclusion that the lower energy fluctuation data are anomalous \cite{NA49 fluctuations 1}.  In these fits,  $\gamma_{q} > 1$ (chemically over-saturated) for $\sqrt{s_{NN}} < 9$ GeV and $\gamma_{q} < 1$ (chemically under-saturated) for $\sqrt{s_{NN}} >$ 9 GeV.

\begin{figure}
	\centering
	\includegraphics[width=3.0 in]{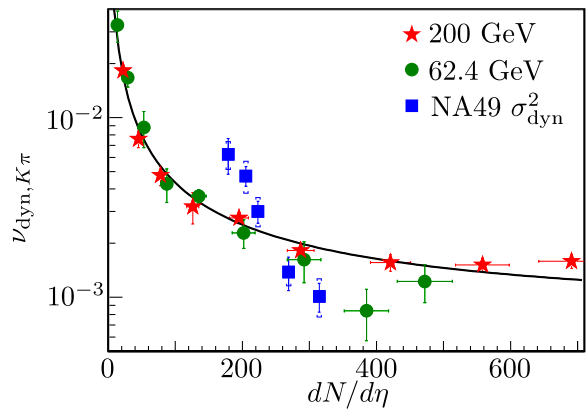}
	\caption{\label{fig:fig3}(Color online) Measured dynamical $K/\pi$ fluctuations in terms of $\nu_{{\rm dyn},K\pi}$ for 62.4 and  200 GeV Au+Au compared with $\sigma_{\rm dyn}^2$ from central Pb+Pb collisions at  6.3, 7.6, 8.8, 12.3, and 17.3 GeV from NA49 \cite{NA49 fluctuations 1}.  Statistical errors are shown for the STAR data.  Statistical and systematic errors are shown for the NA49 results.  The solid line corresponds to a fit to the STAR data of the form $c+d/(dN/d\eta)$.}
\end{figure}

We propose to study $K/\pi$ fluctuations using a variable that does not involve the $K/\pi$ ratio directly.  We choose to employ the variable $\nu_{{\rm dyn},K\pi}$ , which is similar to the observable $\nu_{+-,{\rm dyn}}$ \cite{STAR_net_charge} used to study net charge fluctuations.  $\nu_{{\rm dyn},K\pi}$ does not require mixed events and does not depend on detector efficiencies.  In Fig.~\ref{fig:fig3}, we show  $\nu_{{\rm dyn},K\pi}$ for 62.4 and 200 GeV Au+Au collisions plotted as a function of  $dN/d\eta$.  We estimate the systematic error in $\nu_{{\rm dyn},K\pi}$ to be 15\% due to losses from the electron cut.  Using HIJING \cite{hijing} , we estimate that the effect of feed down on $\nu_{{\rm dyn},K\pi}$ from weakly decaying particles is 9\%.  HIJING calculations show that increasing the accepted range in $p_{t}$ from $0.2 < p_{t} < 0.6$ GeV/c to $0.1 < p_{t} < 2.0$ GeV/c decreases $\nu_{{\rm dyn},K\pi}$ by less than 5\% at both 62.4 and 200 GeV.

In Fig.~\ref{fig:fig3}, we plot the NA49 results using the identity  $\sigma_{\rm dyn}^2 = \nu_{{\rm dyn},K\pi}$.  We verified the validity of this identity explicitly by calculating both quantities for 62.4 and 200 GeV Au+Au collisions.  We find that $\sigma_{\rm dyn}^2 = \nu_{{\rm dyn},K\pi}$ within errors for all centrality bins except the two most peripheral bins at 62 GeV.   In Fig.~\ref{fig:fig3}, we make the correspondence between the incident energy for the NA49 results and $dN/d\eta$ using the systematics in Ref. \cite{phobos}.  The solid line in Fig.~\ref{fig:fig3} represents a fit to the STAR data of the form $c + d/(dN/d\eta)$ where $c$ and $d$ are constants.    The fit for $\nu_{{\rm dyn},K\pi}$ versus $dN/d\eta$ has a $\chi^2$ of 26.6 for 16 degrees of freedom.  If we make a similar fit for $\nu_{{\rm dyn},K\pi}$ versus $N_{\rm part}$, we obtain a $\chi^2$  of 50.7 for 16 degrees of freedom. Thus, the fit for $\nu_{{\rm dyn},K\pi}$ versus $dN/d\eta$ is significantly better than the fit for $\nu_{{\rm dyn},K\pi}$ versus $N_{\rm part}$.  The NA49 results shown in Fig.~\ref{fig:fig3} show a steeper dependence on $dN/d\eta$ than the STAR data and have a $\chi^2$ of 50.8 for five degrees of freedom compared to the best fit to the STAR data.

Using the results of the fit for $dN/d\eta$ along with the systematics in Ref. \cite{phobos} and the relationship $\sigma_{\rm dyn}^2 = \nu_{{\rm dyn},K\pi}$, we can draw the solid line shown in Fig.~\ref{fig:fig2}.  The line relates the incident energy dependence of $\sigma_{\rm dyn}$ in central collisions to the collision centrality dependence of $\nu_{{\rm dyn},K\pi}$ at higher energies.
We observe a slight increase in the systematic behavior of $\sigma_{\rm dyn}$ based on scaled results from STAR as the incident energy is lowered, while the NA49 results show a steeper increase as the energy is lowered.

\begin{figure}[htb]
	\centering
	\includegraphics[width=3.0 in]{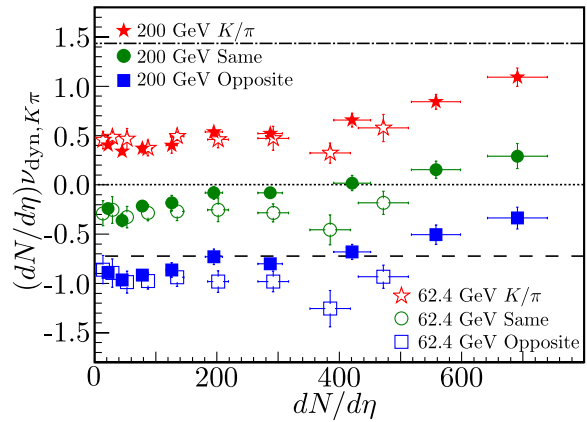}
	\caption{\label{fig:fig4}(Color online) The $dN/d\eta$ scaled dynamical $K/\pi$ fluctuations for summed charges (stars), same signs (circles), and opposite signs (squares) as a function of $dN/d\eta$.  The errors shown are statistical.  The open and filled symbols refer to Au+Au collisions at 62.4 GeV and 200 GeV respectively.  The dash-dot, dotted, and dashed lines represents HIJING calculations for summed charges, same signs, and opposite signs respectively.}
\end{figure}

In order to gain insight into the origin of these $K/\pi$ fluctuations, we calculate $\nu_{{\rm dyn},K\pi}$ for $K^{+}/\pi^{+}$, $K^{-}/\pi^{-}$, $K^{+}/\pi^{-}$, and $K^{-}/\pi^{+}$.  We observe that, within errors, $\nu_{{\rm dyn},K^{+}\pi^{+}}$ is equal to $\nu_{{\rm dyn},K^{-}\pi^{-}}$ and $\nu_{{\rm dyn},K^{+}\pi^{-}}$ is equal to $\nu_{{\rm dyn},K^{-}\pi^{+}}$.  We report the average of the fluctuations of the ratios $K^{+}/\pi^{+}$ and $K^{-}/\pi^{-}$ as same sign and the average of the fluctuations of the ratios $K^{+}/\pi^{-}$ and $K^{-}/\pi^{+}$ as opposite sign in Fig.~\ref{fig:fig4} along with results for $K/\pi$ as a function of $dN/d\eta$. Because $\nu_{{\rm dyn},K\pi}$ scales with the inverse multiplicity, we multiply  our results for $\nu_{{\rm dyn},K\pi}$ by $dN/d\eta$ to study the collision centrality dependence more effectively. 

The scaled  $\nu_{{\rm dyn},K\pi}$ results for all cases in Fig.~\ref{fig:fig4} become more positive as the collisions become more central.  The scaled results for the summed signs are always positive and increase as the collisions become more central.  The scaled results for the opposite sign are always negative indicating a strong correlation for opposite sign particles.  One might expect such negative opposite sign correlations from processes such as the decay $K^* \left( {892} \right) \to K^ +   + \pi ^ -$.  The scaled  $\nu_{{\rm dyn},K\pi}$ for the same sign are slightly negative in peripheral collisions and slightly positive in central collisions, crossing zero around $dN/d\eta$ = 400.  The fact that  $\nu_{{\rm dyn},K\pi}$ for same sign particles is close to zero indicates that the correlations between same sign particles are small.

Also shown in Fig.~\ref{fig:fig4} are filtered HIJING calculations for 62.4 and 200 GeV Au+Au collisions.  These calculations include the acceptance cuts of $|\eta| < 1.0$ and $0.2 < p_{t} < 0.6$ GeV/c. In contrast to the experimental results, HIJING predicts little or no collision centrality dependence for $(dN/d\eta)\nu_{{\rm dyn},K\pi}$.  Because of this lack of centrality dependence, we show the average of the HIJING predictions as a horizontal dot-dashed line.  HIJING predicts that the same sign fluctuations are always zero (dotted line) and the opposite sign fluctuations are always negative (dashed line).
HIJING does not include resonances that decay to $K^{+} + \pi^{+}$ or $K^{-} + \pi^{-}$.
One explanation for the increase in the measured same sign and opposite sign fluctuations scaled with $dN/d\eta$ may be that in peripheral collisions, products of the decay of resonances emerge without further interaction, leading to negative values of $\nu_{{\rm dyn},K\pi}$.  In central collisions, the daughters of the decay of resonances are rescattered in or out of our acceptance, leading to more positive values of $\nu_{{\rm dyn},K\pi}$.  For example, the decay of $K_1 \left( {1270} \right)^+ \to K^+ + \rho^0  \to K^+ + \pi ^ +   + \pi ^ -$ would lead to negative same sign fluctuations in peripheral collisions but not in central collisions.

In conclusion, we find that the fluctuations in the $K/\pi$ ratio for central Au+Au collisions at $\sqrt{s_{NN}}$ = 19.6, 62.4, 130, and 200 GeV are of the same order as the fluctuations observed in central Pb+Pb collisions at  $\sqrt{s_{NN}}$ = 6.3, 7.6, 8.8, 12.3, and 17.3 GeV using the variable $\sigma_{\rm dyn}$, but the Pb+Pb results show a stronger incident energy dependence.  The statistical hadronization model of Ref. \cite{torrieri} cannot reproduce the incident energy dependence of these fluctuations. The collision centrality dependence of $K/\pi$ fluctuations for Au+Au collisions at $\sqrt{s_{NN}}$ = 62.4 and 200 GeV as characterized by the variable $\nu_{{\rm dyn},K\pi}$ seems to scale with $dN/d\eta$. Relating the observed centrality dependence of these fluctuations observed in Au+Au collisions at $\sqrt{s_{NN}}$ = 62.4 and 200 GeV using $\nu_{{\rm dyn},K\pi}$ to the incident energy dependence of fluctuations in central collisions using $\sigma_{\rm dyn}$, we find a smooth scaling, decreasing slightly with increasing incident energy. The scaled values fall in the middle of the fluctuations observed by NA49 between $\sqrt{s_{NN}}$ = 6.3 and 17.3 GeV. $dN/d\eta$ is a good measure of the freeze-out volume and we have shown that the fluctuations in $K/\pi$ scale with $dN/d\eta$.  More measurements are required to demonstrate if there is any discontinuity in $K/\pi$ fluctuations as a function of incident energy.  $\nu_{{\rm dyn},K\pi}$ results using pions and kaons with the same sign are close to zero while results using $\nu_{{\rm dyn},K\pi}$ for opposite sign pions and kaons are negative.  The results for $\nu_{{\rm dyn},K\pi}$ scaled by $dN/d\eta$ become more positive as the collisions become more central due to rescattering, unlike the predictions of HIJING that show no centrality dependence.  These results may indicate that, due to later stage hadronic rescattering, the decay products of resonances are less likely to survive in central collisions than in peripheral collisions.

We thank the RHIC Operations Group and RCF at BNL, and the NERSC Center 
at LBNL and the resources provided by the Open Science Grid consortium 
for their support. This work was supported in part by the Offices of NP 
and HEP within the U.S. DOE Office of Science, the U.S. NSF, the Sloan 
Foundation, the DFG cluster of excellence `Origin and Structure of the Universe', 
CNRS/IN2P3, RA, RPL, and EMN of France, STFC and EPSRC of the United Kingdom, FAPESP 
of Brazil, the Russian Ministry of Sci. and Tech., the NNSFC, CAS, MoST, 
and MoE of China, IRP and GA of the Czech Republic, FOM of the 
Netherlands, DAE, DST, and CSIR of the Government of India,
the Polish State Committee for Scientific Research,  and the Korea Sci. 
\& Eng. Foundation.

\end{document}